\documentclass[aps,prb,preprint,amsmath,amssymb]{revtex4-2}
\usepackage{graphicx}
\usepackage{bm}

\newcommand{\dif}{\mathrm{d}}
\newcommand{\me}{\mathrm{e}}
\newcommand{\mi}{\mathrm{i}}
\newcommand{\bra}[1]{\left\langle #1 \right\rvert}
\newcommand{\ket}[1]{\left\lvert #1 \right\rangle}
\newcommand{\braket}[2]{\left\langle #1 \! \right. \left| #2 \right\rangle}

\begin{document}

\preprint{LA-UR-21-20902}

\title{
\textbf{Quantum Theory of Measurement}
}

\author{Alan K. Harrison}
\email{alanh@lanl.gov}

\affiliation{Los Alamos National Laboratory\\
MS T086, P O Box 1663\\
Los Alamos, New Mexico 87545\\
}

\date{\today}

\begin{abstract}
We describe a measurement in quantum mechanics
as a variational principle
including a simple interaction between the system under measurement and the
measurement apparatus.
Augmenting the action with a nonlocal term (a double integration over the
duration of the measurement interaction) results in a theory capable of
describing both the measurement process (agreement between system
state and the pointer state of the measurement apparatus) and the collapse
of both systems into a single eigenstate (or superposition of degenerate
eigenstates) of the operator corresponding to the measured variable.
In the absence of the measurement interaction, a superposition of states
is stable, and the theory agrees with the predictions of standard quantum
theory. Because the theory is nonlocal, the resulting wave equation is
an integrodifferential equation (IDE).
We demonstrate these ideas using a simple Lagrangian for both systems,
as proof of principle. The variational principle is time--symmetric and
retrocausal, so the
solution for the measurement
process is determined by boundary conditions at both initial and final times;
the initial condition is determined by the experimental preparation
and the final condition is the natural boundary condition of variational
calculus. We hypothesize that one or more hidden variables (not
ruled out by Bell's Theorem, due both to the retrocausality and the
nonlocality of the theory) influence the outcome of the measurement,
and that distributions of the hidden variables that arise plausibly
in a typical ensemble of experimental realizations give rise to
outcome frequencies consistent with Born's rule. We outline
steps in a theoretical validation of the hypothesis.
We discuss the role of both initial and final conditions to determine a solution
at intermediate times, the mechanism by which a system responds to measurement,
time symmetry of the new theory, causality concerns, and issues surrounding
solution of the IDE.
\end{abstract}

\maketitle

%
%
\section{Introduction}
\subsection{Motivation and philosophical stance}
Quantum theory in general, and its description of measurement in particular,
seems to violate several reasonable expectations about the the characteristics
of a correct physical theory. Ordinarily, to be accepted as correct and complete,
a theory must predict future phenomena, given a complete set of the relevant
initial conditions. Quantum theory fails to do this in the case of a measurement;
in fact, it is understood that the mathematical description (wave equation)
describing system
evolution in the absence of a measurement \emph{does not apply} to a
measurement. In effect, two different theories are required, for the
measurement and non--measurement cases. While it may be acceptable
for a theory to treat different cases in different ways, quantum theory lacks
an unambiguous definition of a measurement, with the result that
measurement and non--measurement configurations 
may be arbitrarily similar physically, and the bipartite theoretical description
is implausible.

In addition, the theory of quantum measurement (as distinguished from
the wave equation) as usually interpreted (e.g. by the Copenhagen
interpretation) has multiple features that are unknown in any other
generally accepted fundamental theory. One is intrinsic randomness, the idea that
nature samples from a random distribution, and no prediction
can be made about the result of sampling that goes beyond
a description of the distribution function.
Another is temporal asymmetry;\footnote{
We are aware of course that thermodynamics seems to have a preferred direction
of time, but point out
that the fundamental dynamic laws that give rise to it are time--symmetric.
}
after the measurement, but not before,
the system is understood to be ``collapsed'' into an eigenstate or set of degenerate
eigenstates of the operator corresponding to the measured quantity.

A third feature unique to the quantum measurement process is dependence
on the eigenstate structure of the problem.
The observed behavior that a measurement always
finds the system in a single eigenstate (or a superposition of degenerate
eigenstates) of the operator
requires a nonlocal theory. As we will discuss in subsection \ref{Necessity_nonlocality},
the information (e.g., potential $V(x_0)$) available at a single point $x_0$
is insufficient to determine whether a particular solution at that point
(values of the wavefunction $\psi(x_0)$ and its derivative(s) at $x_0$) is
consistent with a \emph{single eigenfunction} $\psi$ (the function defined for
all allowed values of $x$). Nature cannot reliably make that determination
at $x_0$ without using information at points $x \neq x_0$.

In addition, we call attention to quantum phenomena that seem to violate causality.
One is correlations
between spacelike separated measurements in ways that violate
special--relativity--based expectations (``EPR correlations,'' for short)
\cite{EPR} and Bell's inequality \cite{Bell_1964}
but have been verified in a long sequence of increasingly more
sophisticated experiments
\cite{CHSH, *Freedman_Clauser, *Aspect_Grangier_Roger, *Aspect_Dalibard_Roger}.
Another is delayed--choice experiments \cite{Wheeler_1979}, in which the path of a
particle (through one or two slits, for example) has been observed to be apparently
determined by a choice made after the particle is committed to a particular path.

In this paper, we propose that a quantum theory can be constructed so as to either
avoid or explain
most of these objectionable or unique features. To be specific, we will exhibit a wave
equation that applies even when a measurement is being done, in which case it
describes evolution (``collapse,'' although not instantaneous) of the wavefunction
to a state or states with a single eigenvalue. The theory is time--symmetric.
Instead of relying on intrinsic randomness to explain differing results of
identically prepared measurements, it proposes that some hidden variable(s),
presumably uncontrolled or overlooked by the experimenter, determine(s)
the outcome. The Born's--rule distribution of outcomes \cite{Born} attributed to randomness
by standard quantum theory presumably appears instead as a result of a
naturally--arising distribution of values of the hidden variable(s)---although
the complete proof of that result must await further investigation.

On the other hand, the theory we describe relies on some unusual assumptions;
we do not expect to replace conventional quantum theory with one that
completely resembles other physical theories.
One such assumption is retrocausality, roughly speaking, the idea that
effects may precede their causes in time. (To be more precise, in a
retrocausal theory the solution at $t$ is found as a function of variables
at $t'>t$.) The theory is also nonlocal,
as needed to produce the needed dependence on eigenstate structure;
for this reason the wave equation is a integrodifferential equation (IDE).
Finally, as mentioned earlier, we posit
the existence of hidden variables. Bell's Theorem and its experimental
tests are generally understood to rule out local hidden--variable theories,
but that does not restrict our nonlocal theory. In addition, it has been
pointed out \cite{Argaman_2010} that the proof of Bell's theorem relies on
an assumption violated by retrocausality, so for that reason also, hidden
variables are not off limits in this case.

We point out here that because retrocausality can allow information to
propagate backward in time, it trivially explains EPR correlations and
delayed--choice experiments.\cite{Sutherland_2017}. Since those two issues
are already disposed of, we will focus our attention on the remaining ones.

\subsection{Elements of the theory}
We consider that a legitimate measurement is understood to require a duration
$T$ limited by the time-energy uncertainty relation
\begin{equation}
\label{t-E_uncertainty}
T \, \Delta E \ge \hbar/2
\end{equation}
where $\Delta E$ is the smallest energy difference between states that must be
distinguished by the measurement. Typical experiments are designed with
$T \, \Delta E \gg \hbar/2$. In our analysis, we will suppose that the system is
prepared and the experiment begun at time $t_i$, and the measurement is
determined or read at $t_f = t_i +T$.

We desire the theory to be time--symmetric, and it is appealing to do so
by couching it as a variational principle. In this case the state
$\psi$ of the system is found to be a critical point of the action
\begin{equation}
I[\psi] \equiv \int_{t_i}^{t_f} L[\psi,\dot{\psi},t]
\end{equation}
where $L$ is the system Lagrangian, typically the spatial integral of a
Lagrangian density. Critical points are choices of the function $\psi$ where $I$
is stationary with respect to infinitesimal variations of $\psi$.
For functionals that depend smoothly on their arguments, maxima and minima are
critical points, so the search for critical points is often described as finding extrema.
Schwinger \cite{Schwinger_1951} developed quantum field theory
as a variational principle
based on the action. For our purposes,
we note that the action has the same value regardless of the direction of time,
so the resulting theory is time--reversal invariant.

Our exposition will be nonrelativistic,
but the variational principle is inherently compatible
with special relativity \cite{Schwinger_1951},
and we expect that it can readily be expressed in a relativistically covariant
formulation. Relativistic Lagrangians routinely appear in quantum field
theory,\cite{Weinberg_1995} and the four-dimensional integration of the Lagrangian density
to produce the action is of course a relativistically appropriate operation,
invariant under change of reference frames.

In its simplest form, a variational principle leads to a differential equation, the
Euler equation \cite{Courant}. In order to introduce nonlocality, we will employ
a more complex form (a double rather than a single integral in time) that 
will result in an integrodifferential equation (IDE); see Appendix \ref{Two_time_variant}
for the mathematical details. The IDE involves an integral from $t_i$ to $t_f$,
so nonlocality in time is evident. Note that the conventionally--understood unitary
evolution of the system (as described by the wave equation) would be predicted
by the conventionally--derived action, without our modification. We expect that
the modified action will predict, via the variational principle, the combination of the
non--measurement evolution of the system and the effect of the measurement.

This mathematical form apparently requires solving for $\psi$ simultaneously for all times
in $[t_i, t_f]$. This contrasts with a typical physical theory in which variables and their
time derivatives at $t$ depend on other variables and derivatives at $t$, or in some cases
on $t$ and its past. Wharton \cite{Wharton_2012} has designated
these two approaches as Lagrangian and Newtonian respectively, and argued
persuasively that the former may be appropriate for physical theories.
Note that this picture is definitely retrocausal, because the wavefunction at time
$t$ may depend on conditions or the wavefunction at times $>t$, and in
particular at $t_f$.

The most obvious way to solve such a mathematical problem is with specified
initial and final conditions $\psi(t_i)$ and $\psi(t_f)$. Consider a typical measurement
problem in which the system to be measured is prepared at $t_i$ in a given quantum
state, defined as an eigenstate or a specified superposition of eigenstates of
a given operator. Then a measurement concluding (``read out'') at $t_f$
determines in which of the eigenstates of that operator the system is found
at that time. In this case the initial condition is fixed by the specified experimental preparation,
but the final condition appears to be missing. Calculus of variations \cite{Courant}
supplies the missing constraint, namely, a ``natural boundary condition'' (NBC) that
inevitably applies at a boundary where the value of the unknown function is
not specified by the problem definition.

As a familiar example, consider a vibrating string of length $L$. It is described by a simple
wave equation (a differential equation) for the displacement $y(x,t)$,
but we could
equally well cast the problem as a variational principle and deduce the wave
equation from the system Lagrangian. Now if the string is fixed at both ends,
the variational principle (and hence the wave equation) must satisfy ordinary
BCs at both ends: $y(0,t)=y(L,t)=0$. But if at $x=0$ the string
is not fixed but rather free to slide frictionlessly along a rod perpendicular to
the string, the BC at that point is the NBC
$\partial y/\partial x \rvert_{x=0} = 0$. Note that that condition is caused not by
the rod, which does not constrain the string's position or slope,
but by the Lagrangian, by which the condition follows from the requirement of
stationarity of the action.

We conclude that the solution of the IDE must be 
constrained by the specified preparation at $t_i$ (an initial condition) and the
NBC at $t_f$ (a final condition). However, the empirical fact that different
outcomes may result from identically--prepared repetitions of the same
measurement proves that those two conditions underconstrain the problem.
At $t_i$, the system is prepared in a given quantum state or superposition
of states (e.g., the ground state of a square--well potential), but that description
falls short of a specification of every possible
variable (including e.g. both position and momentum),
as it must by quantum complementarity
\cite{Wharton_2010, *Wharton_2010a}. The full specification of the initial
(ontological) state consists of the given quantum state, plus additional
``hidden variables'' unknown to or uncontrolled by the experimenter.
Similarly, the measurement of a quantum state at $t_f$ does not determine
the ontological state at that moment; in fact, the measurement readout at
$t_f$ is a weaker constraint than the preparation at $t_i$, because it
determines only the variable (operator) measured but not its value
(eigenvalue).\footnote{The asymmetry between $t_i$ and $t_f$ arises
from the measurement process we have described; the general theory
is still time--symmetric.}
This indeterminacy provides the opportunity for hidden variables to
participate in determining the result of the measurement.

We will see that our theory predicts the collapse
(not necessarily instantaneous; perhaps a better term is decay) of the wavefunction to a
single eigenvalue at $t_f$. We expect that the BCs together with the
hidden variable(s) determine which final state results from the collapse.
Ultimately, the frequencies of the different outcomes possible from a single
experimental definition must reflect the distribution of hidden variable values
in a large number of realizations of the experiment. The observed fact that
that those frequencies may be described by a simple law (Born's rule)
presumably reflects a likelihood of an approximately universal distribution
of the hidden variable values in experiments that are likely to be conducted
(without knowledge or control of the hidden variable(s)). 
For instance, suppose the experimental result depends on a high--frequency
sinusoidal
function of some experimental time. If in an ensemble of experimental
realizations that time is naturally distributed over a range large compared
to the period of oscillation, it is an excellent approximation to say that that time
has a uniform distribution over a single period. In this way, it is reasonable to
expect that naturally--occurring
ensembles of experiments may be found reliably to give outcome
frequencies satisfying Born's rule.

\subsection{Model of the measurement problem}
As the principal issues motivating our theoretical development have to do
with quantum measurement, we will consider an idealized model of such a
measurement. 
Suppose that the system is prepared in a known superposition 
$\sum_j C_j \ket{\psi_j}$
of eigenstates
of the operator $\sigma_{op}$ at time $t_i$; that is, the eigenstates are
well--defined and the coefficients $C_j(t_i)$ are known. This superposition
is known to be initially stable; $\dot{C}_j(t_i)=0$ for all $j$.
The eigenstates themselves must be stable, so $\sigma_{op}$ commutes
with the Hamiltonian.
Finally, the stability of an (unperturbed, unmeasured) superposition 
implies that the system is linear when in isolation, that is,
it satisfies a wave equation linear
in the wavefunction.
This consideration will be seen to constrain the form of possible Lagrangians
for the system. We will develop our ideas using a particular simple form,
as proof of principle.

During all or part of the interval
$[t_i,t_f]$, a measurement apparatus (which we will call system 2)
interacts with the measured system (system 1).
A requirement for generality of the theory---validity
of the properties of ``quantum measurements'' 
across all types of measurements---excludes all but the most general description of the 
measurement apparatus and its interaction with the measured
system. We therefore use a minimal
description, that the apparatus has a ``pointer state'' variable $\sigma^2$, 
and that it is coupled to
the measured variable $\sigma^1$ of the system.
Without loss of generality, we define $\sigma^2$
so that its value in a successful measurement equals the value of $\sigma^1$.
Then the composite (system $+$ apparatus) Lagrangian must include an interaction term
that depends on both measured and pointer state variables, and attains an extreme (or
stationary) value when they are equal. The simplest such term is quadratic in the corresponding
operators, that is, proportional to
$(\sigma_{op}^1 - \sigma_{op}^2)^2$.

Note that good experimental design dictates that the combined system
(1 and 2) be well isolated in spacetime. Spatial isolation is accomplished
by physical isolation or other control of the boundaries of the domain,
and temporal isolation by system preparation at $t_i$ and measurement
readout at $t_f$. This blocks influences from outside the spacetime
region, which is important so that the spacetime integrals in this
nonlocal theory can legitimately be limited to the experimental domain.

What is known experimentally is that
if there is no measurement, system 1 remains indefinitely in the same superposition
of states in which it was prepared.
If there is a measurement, it is found (measured) to be in a single eigenstate.
(Actually, this may be a superposition
of degenerate eigenstates---states with a single eigenvalue.)
Finally, in an ensemble of identically prepared measurements,
measured eigenstates occur in proportion to their
weight $\lvert C_j(t_i) \rvert^2$ in the initial superposition (Born's rule).
We seek a theory that predicts these empirical facts.

This description of the ``measurement problem'' is understood to be very well established
by a large body of experimental evidence. On the other hand, that body of evidence is 
silent on the outcomes of measurements violating (\ref{t-E_uncertainty}),
because such experiments
would be understood to be ineligible to invalidate any of the above points.
In other words, we may consider Born's rule to be a summary of observations
about experiments conforming
to (\ref{t-E_uncertainty}), since nonconforming experiments would not have been considered proper
measurements.

\subsection{Outline}
In the next section, we will develop the theory based on
a variational principle, generalized so as to result in a nonlocal equation.
The subsequent section will discuss the predictions of that equation and
compare them to the properties that we have argued must appear in
a successful theory. In some cases the agreement will be clear,
although it will remain for the future to describe the details of
approach to the solution, and to prove that the solution is unique.
For Born's rule, we will show how hidden variables may arise
and the way in which the expected output frequencies may follow
from their distribution; however, analytic proof or numerical
demonstration that our theory yields frequencies consistent
with Born's rule remains to be done. In the last section we will summarize
what we have done, discuss new perspectives required by
retrocausality and nonlocality, and list some of the next steps
to be taken to continue developing these ideas. A mathematical
appendix derives an extension of variational calculus used in
our analysis of the nonlocal variational principle.

\section{Theoretical development}
\label{Theoretical_development}

\subsection{Variational approach}
\label{Variational_approach}
An isolated system (system 1 or system 2, in our case, when they are not interacting),
is described by a Lagrangian $L[\psi,\dot{\psi},t]$---a functional of the wavefunction $\psi(t)$---which is
typically the spatial integral of a Lagrangian density $\mathcal{L}[\psi(t,\boldsymbol{x})]$.
The variational principle (to be specific, the Hamiltonian principle) says
that the action $S \equiv \int \dif t L$ is stationary with respect to variations of $\psi$,
a condition that may be denoted $\delta S = 0$. 
(This is the principle that was employed by Schwinger\cite{Schwinger_1951}
as the foundation for quantum field theory.)
A choice of $\psi$ for which $S$ is stationary is said to be a critical point of $S$.
For functionals that depend smoothly on their arguments, maxima and minima are
critical points, so the search for critical points is often described as finding extrema.
A necessary condition is given by
the Euler equation
\begin{equation}
\label{Euler}
0 = \frac{\partial L}{\partial \psi}
 - \frac{\dif}{\dif t} \frac{\partial L}{\partial \dot{\psi}}
\end{equation}
Evidently the requirement that (\ref{Euler}) yield a linear wave equation
implies that the Lagrangian must be quadratic in $\psi$ and its time derivative.

\subsection{Normal--mode expansion---Single system}
Since the point of the measurement problem is to describe the evolution of a superposition of
eigenstates of a given operator to a single eigenstate, it will simplify matters to define a basis
set of such eigenstates. This expansion will be specific to a given inertial reference
frame---the frame in which the measurement is performed and described by the above
characteristics---because that will simplify the analysis and its comparison to those points. 
However, as explained above, we expect that the general theory
(the form of the action, without dependence on
the normal--mode expansion we will use here) will be relativistically appropriate and can be
expressed in covariant form.

We will describe each system $\ell$ (=1 or 2) by a wavefunction $\psi^\ell(t,\boldsymbol{x})$, normalized
in the usual way in terms of the spatial integral or otherwise--defined inner product
\begin{equation}
\label{normalization_psi}
\braket{\psi^\ell}{\psi^\ell} = 1
\end{equation}

At any given time $t$, let $\ket{\psi^\ell_j}$ be for system $\ell = 1$ or 2 an eigenstate
of a Hermitian operator $\sigma_{op}^\ell$,

\begin{equation}
\label{eigenvalue}
\sigma_{op}^\ell \ket{\psi^\ell_j(t)} = \sigma_j^\ell \ket{\psi^\ell_j(t)}
\end{equation}

\noindent
satisfying the applicable spatial BCs, and let those eigenstates form an
orthonormal basis for states of system $\ell$:

\begin{equation}
\label{orthonormality}
\braket{\psi^\ell_j(t)}{\psi^\ell_k(t)} = \delta_{jk} \qquad (\ell=1,2; \forall t)
\end{equation}

Since external fields acting on the system may change during the course of the measurement
(perhaps due to the measurement process itself), the eigenvalues and eigenstates are in general
functions of time. In many interesting cases
they are slowly varying functions of time, and for simplicity we will confine ourselves
to the case in which the eigenvalues $\sigma_j^\ell$ are constant. We expect that the
analysis presented below can be readily generalized to the time--dependent case, for
sufficiently slow variation.

We will also require each normal mode $\ket{\psi^\ell_j}$ to satisfy the variational principle based on its
single-system Lagrangian $L^\ell$. This is possible because as stated above, the operator corresponding
to the measured variable commutes with the Hamiltonian. The basis states will be taken to be
simultaneous eigenstates of both operators, and eigenstates of the Hamiltonian satisfy the
variational principle.
Since a basis vector $\ket{\psi_j^\ell(t)}$ was defined to be an eigenstate
of the Hamiltonian, it has an energy $E_j^\ell$
and a time derivative
\begin{equation}
\label{energy}
\frac{\dif}{\dif t} \ket{\psi_j^\ell(t)} =
-\frac{\mi}{\hbar}E_j^\ell \ket{\psi_j^\ell(t)}
\end{equation}
(Sch\"{o}dinger picture). We will also take the energies $E_j^\ell$ to be constant; then it follows that
\begin{equation}
\label{modal_inner_product}
\braket{\psi^\ell_j(t_1)}{\psi^\ell_k(t_2)} = \delta_{jk} \,
\me^{-\frac{\mi}{\hbar} E_j^\ell(t_2-t_1)}
\end{equation}

Now if system $\ell = 1$ (measured system) or $2$ (measurement apparatus) is isolated,
its wavefunction can be expanded
\begin{equation}
\label{C_j_expansion}
\ket{\psi^\ell(t)} = \sum_j C^\ell_j(t) \ket{\psi^\ell_j(t)}
\end{equation}
and the normalization condition (\ref{normalization_psi}) implies
\begin{equation}
\label{normalization_Cj}
\sum_j \lvert C_j(t) \rvert^2 = 1
\end{equation}
At present we expect this condition to hold for any $t$, but in subsection
\ref{Alternative_normalization} we will argue for removing this constraint.

The action is
\begin{eqnarray}
\label{action_single}
S^\ell &\equiv& \int_{t_i}^{t_f} \dif t \, L^\ell(t)
\nonumber\\
&=& \int_{t_i}^{t_f} \dif t \bra{\psi^\ell(t)} L_{op}^\ell \ket{\psi^\ell(t)}
\nonumber\\
&=& \sum_{j,k} \int_{t_i}^{t_f} \dif t \bra{\psi^\ell_j(t)} \, C^{\ell*}_j(t) \, L_{op}^\ell \, C^\ell_k(t) \ket{\psi^\ell_k(t)}
\end{eqnarray}

Since the complete wavefunction $\psi^\ell$ is completely determined by the
set of coefficients $C_j^\ell(t)$, the condition of stationarity of the action
reduces to the problem of finding those coefficients, which must satisfy
\begin{equation}
\label{E-L_Cjk}
0 = \frac{\partial L^\ell}{\partial C^\ell_j}
 - \frac{\dif}{\dif t} \frac{\partial L^\ell}{\partial \dot{C}^\ell_j}
 \quad \forall j
\end{equation}
This formulation of the problem replaces (\ref{Euler}). 

It is traditional in quantum field theory to perform the variational calculus analysis
by varying (differentiating with respect to)
the physically significant canonical fields and momenta, and that approach
is extremely useful in producing intuitively appealing and useful
evolution equations.\cite{Schwinger_1951, Weinberg_1995}
However, the stationarity
of the action is a \emph{mathematical} condition, and as long as our formulation
spans the space of its allowed variations, the mathematics
does not dictate our choice of the functions in terms of which those variations
are expressed. Because we are interested in the eigenstate
content of the wavefunction, the corresponding coefficients are particularly
useful to us, and we use them to analyze the variational principle.

\subsection{Combined systems}
Now we can use the normalization condition (\ref{normalization_psi}) to write
from (\ref{action_single})
\begin{equation}
\label{combined_action_no_interaction_alt}
S^1 + S^2 = 
\int_{t_i}^{t_f} \dif t
\bra{\psi^1(t)} \bra{\psi^2(t)} 
(L_{op}^1 + L_{op}^2)
\ket{\psi^1(t)} \ket{\psi^2(t)}
\end{equation}
if there is no interaction or entanglement between the two systems,
that is, the combined state factors as $\ket{\psi} \equiv \ket{\psi^1}\ket{\psi^2}$.

To allow the two subsystems to be entangled, we replace the product of single-system states $\ket{\psi^1}$ and $\ket{\psi^2}$
by the joint state
\begin{equation}
\label{C_jk_expansion_alt}
\ket{\psi(t)} = \sum_{j,k} C_{jk}(t) \ket{\psi^1_j(t)} \ket{\psi^2_k(t)}
\end{equation}
whereupon the normalization condition becomes
\begin{equation}  
\label{normalization_Cjk}
\sum_{j,k} \, \lvert C_{jk}(t) \rvert^2 = 1 \quad \forall t
\end{equation}
Then
\begin{equation}
\label{action_single_system_terms}
S^1 + S^2
= \sum_{j,k,\ell,m}
\int_{t_i}^{t_f} \dif t \,\,
\bra{\psi_j^1(t)} \, \bra{\psi_k^2(t)} \, C^*_{jk}(t)
(L_{op}^1 + L_{op}^2)
\, C_{\ell m}(t) \, \ket{\psi_\ell^1(t)} \, \ket{\psi_m^2(t)}
\end{equation}
and

To simplify the single-system terms, suppose $L^1_{op}$ and $L^2_{op}$ are of the form
\begin{eqnarray}
\label{L_op_form}
L^\ell_{op}
&=& A^\ell - B^\ell \, \frac{\dif^2}{\dif t^2}
\nonumber\\
&=& A^\ell + \overleftarrow{\frac{\dif}{\dif t}} \, B^\ell \, \frac{\dif}{\dif t}
\end{eqnarray}
so $L^1$ and $L^2$ take the form
\begin{equation}
\label{Ll_form}
L^\ell \equiv \bra{\psi^\ell} L^\ell_{op} \ket{\psi^\ell}
= A^\ell \braket{\psi^\ell}{\psi^\ell} + B^\ell \braket{\dot{\psi}^\ell}{\dot{\psi}^\ell}
\end{equation}
with real constants $A^\ell$ and $B^\ell$.
Then the fact that $\ket{\psi_j^\ell}$ is an eigenstate of the Hamiltonian means
that it satisfies the Euler equation (\ref{Euler}), which we can write as
\begin{eqnarray}
\label{E-L_psi}
0 &=& \frac{\partial L^\ell}{\partial \bra{\psi_j^\ell}}
 - \frac{\dif}{\dif t} \frac{\partial L^\ell}{\partial \bra{\dot{\psi}_j^\ell}}
 \nonumber\\
&=& A^\ell \ket{\psi_j^\ell} - B^\ell \ket{\ddot{\psi}_j^\ell}
\end{eqnarray}

At this point we observe that the functional
in question is a physical action and therefore real, so it is unchanged if we
drop any imaginary part of the integrand. This has a simplifying advantage.
When we use variational calculus to find a stationary state with respect to variations
of a complex quantity ($\ket{\psi_j^\ell}$ or $C_{jk}$), we may treat the real and
imaginary parts of that quantity independently, with an Euler equation
for each of them. Alternatively, we may treat the quantity and its complex
conjugate ($\bra{\psi_j^\ell}$ or $C^*_{jk}$) as the two functions to be varied.
In our case, with a real integrand, doing so has the
convenient feature that the two resulting Euler equations are complex
conjugates of each other and we only need to solve one of them. Here in
(\ref{E-L_psi}) we choose to vary the bra vector.

Substituting (\ref{L_op_form}) into (\ref{action_single_system_terms}) and
using property (\ref{E-L_psi}) of the eigenvectors, we find (introducing
the shorthand notation
$B \equiv B^1 + B^2$) that
\begin{eqnarray}
S^1 + S^2
&=& \sum_{j,k,\ell,m}
\int_{t_i}^{t_f} \dif t \,\,
\bra{\psi_j^1(t)} \, \bra{\psi_k^2(t)} \, C^*_{jk}(t)
\left[(L_{op}^1 + L_{op}^2),
\, C_{\ell m}(t) \right] \, 
\ket{\psi_\ell^1(t)} \, \ket{\psi_m^2(t)}
\nonumber\\
&=& -B \sum_{j,k,\ell,m}
\int_{t_i}^{t_f} \dif t \,\,
\bra{\psi_j^1} \, \bra{\psi_k^2} \, C^*_{jk}
\left( \ddot{C}_{\ell m} + 2 \, \dot{C}_{\ell m} \frac{\dif}{\dif t} \right) \, 
\ket{\psi_\ell^1} \, \ket{\psi_m^2}
\end{eqnarray}
Then
\begin{eqnarray}
S^1 + S^2
&=& \, -B \! \sum_{j,k,\ell,m}
\int_{t_i}^{t_f} \dif t \,\,
\bra{\psi_j^1} \, \bra{\psi_k^2} \, C^*_{jk}
\left( \ddot{C}_{\ell m} - \frac{2\mi}{\hbar} E_{\ell m} \, \dot{C}_{\ell m} \right) \, 
\ket{\psi_\ell^1} \, \ket{\psi_m^2}
\nonumber\\
&=& \, -B \sum_{j,k}
\int_{t_i}^{t_f} \dif t \,\,
C^*_{jk} \,
\left( \ddot{C}_{jk} - \frac{2\mi}{\hbar} E_{jk} \, \dot{C}_{jk} \right)
\nonumber\\
&=& \, B \sum_{j,k}
\int_{t_i}^{t_f} \dif t \,
\left( \lvert \dot{C}_{jk} \rvert^2 + \frac{2\mi}{\hbar} E_{jk} \, C^*_{jk} \,\dot{C}_{jk} \right)
\end{eqnarray}
where in the last step we rely on the hypothesis that $\dot{C}_{jk}$ vanishes at $t_i$
as a condition imposed by the experimental
preparation, and at $t_f$ since that is implied by the NBC.
Finally, as intended, we discard the imaginary part of the integrand:
\begin{equation}
\label{action_single_system_terms_simplified}
S^1 + S^2
= \, B \sum_{j,k}
\int_{t_i}^{t_f} \dif t \,
\left( \lvert \dot{C}_{jk} \rvert^2
+ \mathrm{Re} \left \{\frac{2\mi}{\hbar} E_{jk} \, C^*_{jk} \,\dot{C}_{jk} \right\} \right)
\end{equation}

\subsection{Interaction term}
As argued above,
we must account for interaction by including in the action
a term proportional to
$(\sigma_{op}^1 - \sigma_{op}^2)^2$.
A simple form for such an interaction
term is
\begin{equation}
\label{simple_interaction}
S^I = \mu \int^{t_f}_{t_i}\dif t \,
\bra{\psi(t)} (\sigma_{op}^1 - \sigma_{op}^2)^2 \ket{\psi(t)}
\end{equation}
for some constant $\mu$.
Then, defining another shorthand notation
$\Delta_{jk} \equiv \sigma_j^1 - \sigma_k^2$,
\begin{eqnarray}
\label{interaction_local}
S^I &=& \, \mu \sum_{j,k,\ell,m}
\int^{t_f}_{t_i} \dif t \,
\bra{\psi_j^1(t)} \, \bra{\psi_k^2(t)} \, C^*_{jk}(t)
\, (\sigma_{op}^1 - \sigma_{op}^2)^2 \,
C_{\ell m}(t) \, \ket{\psi_\ell^1(t)} \, \ket{\psi_m^2(t)}
\nonumber\\
&=& \, \mu \sum_{j,k}
\int^{t_f}_{t_i} \dif t \,
\, \Delta_{jk}^2 \, \lvert C_{jk}(t) \rvert ^2
\end{eqnarray}
Then we might expect the complete action to be
\begin{equation}
S = S^1 + S^2 + S^I
\end{equation}

\subsection{Necessity of a nonlocal theory}
\label{Necessity_nonlocality}
However, as we indicated earlier,
to reproduce the observed behavior that a measurement always
finds the system in a single eigenstate (or a superposition of degenerate
eigenstates) of the operator corresponding to the measured quantity,
the theory must be nonlocal. For a simple example of this, consider
a system described by the one-dimensional Schr\"{o}dinger equation
\begin{equation}
\label{Schrodinger}
\left[ -\frac{\hbar^2}{2m} \frac{\dif^2}{\dif x^2} - V(x) \right] \psi(x) = E \, \psi(x)
\end{equation}
with potential function $V(x)$ and boundary conditions at positions
$x_1, x_2$ (which may be $\pm\infty$). It is customary to require
the solution to be normalized according to
\begin{equation}
\label{simple_normalization}
\int \lvert\psi\rvert^2 \dif x = 1
\end{equation}
although for our purposes it suffices to require that integral to be finite.
Consider the case in
which the potential is attractive and the spectrum of energy eigenvalues $E$ is discrete,
with (for simplicity) no degenerate eigenstates.

Now suppose that the values $\psi(x_0)$ and $\psi'(x_0)$ are proposed as solutions
at some point $x=x_0$, and we ask whether they belong to a solution that is a single
eigenstate.
In a conventional interpretation of quantum mechanics,
this is the question Nature must answer when those values have developed due to
the operation of the wave equation and then a measurement is made, requiring a
single eigenvalue as its result. 
(We are here dealing with the case in which the
measured quantity is energy, but that case is enough to prove our point.)
Nature must decide whether to
accept the proposed values of $\psi(x_0)$ and $\psi'(x_0)$ as given or ``collapse''
to different values consistent with a single eigenstate.

In a local theory, that question must be answered on the basis of local information
alone, that is, $V(x_0)$. That information is insufficient.
With nonlocal information, namely, the entire function $V(x)$, 
it would be possible, given $E$, to find the
solution $\psi(x)$ of (\ref{Schrodinger}) 
by integrating the differential equation
twice. However, for most values of $E$, either the integrated solution violates the
boundary conditions or the normalization integral diverges (or both).
We conclude that determining whether $\psi(x_0)$ and $\psi'(x_0)$ are consistent
with a single eigenstate requires the use of information $V(x)$ at all $x$ to integrate
the solution and test boundary conditions and normalizability.
A local theory cannot make that determination.

Therefore, since a measurement always finds an eigenstate of the relevant operator,
we conclude that its complete mathematical description must be
nonlocal in space. But a description that is
nonlocal in space in one inertial reference frame is nonlocal in both space and time
in any other frame, so in general the description of a measurement must be nonlocal in time
as well.

A differential equation (aside from the specification of
BCs) is local, depending on a function and its derivatives at a single point.
By contrast, a nonlocal relationship is naturally expressed as an integral equation.
Calculus of variations shows us that the stationary states of an integral expression like
$\int du \, F(g(u),g(\dot{u}),u)$ satisfy a differential equation (the Euler equation)
for $g(u)$, so such a description corresponds to a strictly local process. In order to obtain
an integral equation as the simplest description of a measurement process, we need
the action to involve at least \emph{two} integrations of some function of the quantum state.

\subsection{Nonlocal interaction term}
Since the phenomenon that requires nonlocality (measurement--induced collapse
of the wavefunction) is due to the interaction between systems 1 and 2,
we suppose that it is the interaction term $S^I$ that must be made nonlocal.
We propose to add to it a nonlocal piece involving two integrations on time.
We start with an expression resembling $S^I$ in
(\ref{simple_interaction}) but with two integrations on time:
\begin{eqnarray}
&&
\nu
\left[
\int^{t_f}_{t_i}\dif t \,
\bra{\psi(t)}
(\sigma_{op}^1- \sigma_{op}^2)
\ket{\psi(t)}
\right]^2
\nonumber\\
&&
=
\nu \int^{t_f}_{t_i}\dif t_1 \, \int^{t_f}_{t_i}\dif t_2 \,
\bra{\psi(t_1)} \bra{\psi(t_2)}'
(\sigma_{op}^1- \sigma_{op}^2) \, (\sigma_{op}^{1'}- \sigma_{op}^{2'})
\ket{\psi(t_1)} \ket{\psi(t_2)}'
\end{eqnarray}
Here $\nu$ is a real constant,
and the primed $\sigma^\ell_{op}$ operators
combine with the primed bra and ket vectors in an inner product,
as do the unprimed operators and bra and ket vectors.
Now we make changes so as to couple the $t_1$ and the $t_2$ integrals.
We move one of the primes in the operator kernel, changing it from
$(\sigma_{op}^1- \sigma_{op}^2) \, (\sigma_{op}^{1'}- \sigma_{op}^{2'})$
to $(\sigma_{op}^1- \sigma_{op}^{2'}) \, (\sigma_{op}^{1'}- \sigma_{op}^2)$.
We also move the prime from one ket vector to the other.
Finally, we observe that in this form the interaction between the state at
$t_1$ and at that at $t_2$
is independent of the time difference. It may be that that effect weakens with
temporal separation, so a dimensionless non-negative real function
$f(t_1 - t_2)$ 
should be included in the integrand. By symmetry, $f$ must be an even
function, and we expect it to be a monotonically decreasing function of
the absolute value of its argument. For later convenience, let us suppose
that there is a real constant $\tau$ such that $f(t_1 - t_2)=0$
whenever $\lvert t_1 - t_2 \rvert \ge \tau$.
These changes result in the term
\begin{equation}
\label{first_nonlinear_interaction}
R^I \equiv
\nu \int^{t_f}_{t_i}\dif t_1 \, \int^{t_f}_{t_i}\dif t_2 \,
f(t_1 - t_2)
\bra{\psi(t_1)} \bra{\psi(t_2)}'
(\sigma_{op}^1- \sigma_{op}^{2'}) \, (\sigma_{op}^{1'}- \sigma_{op}^2)
\ket{\psi(t_1)}' \ket{\psi(t_2)}
\end{equation}
Physically this expresses an interaction or ``auto-entanglement'' between the state
$\ket{\psi}$ at time $t_1$ and the same state at $t_2$; this is an expression of
retrocausality in the sense that the state at the later time interacts with its
earlier value. A more speculative interpretation, based on the time symmetry of
the variational principle, is that this term
describes interaction between ``forwards'' and ``backwards'' histories.
This sounds very much like the ``transaction'' in Cramer's transactional
interpretation,\cite{Cramer_1980, *Cramer_1986}
but it is not quite the same; Cramer proposed a two--way interaction between lightlike
separated events, whereas our form allows for the possibility of
timelike, lightlike and spacelike interactions. (We may of course restrict those
options as we gain future understanding.)

We point out that for the extreme choice of $f$
\begin{equation}
f(t_1 - t_2) = \delta(t_1 - t_2)
\end{equation}
the integrand takes a more intuitive form in terms of quantum expectation values
$\left\langle \mathcal{O} \right\rangle \equiv \bra{\psi} \mathcal{O} \ket{\psi}$:
\begin{eqnarray}
\bra{\psi(t)} \bra{\psi(t)}'
(\sigma_{op}^1- \sigma_{op}^{2'}) \, (\sigma_{op}^{1'}- \sigma_{op}^2)
\ket{\psi(t)} \ket{\psi(t)}'
&=&
\left\langle \sigma^1 \right\rangle ^2
-2 \left\langle \sigma^1 \sigma^2 \right\rangle
+ \left\langle \sigma^2 \right\rangle ^2
\nonumber\\
&=&
\left\langle (\sigma^1 - \sigma^2)^2 \right\rangle
- \left\langle (\Delta\sigma^1)^2 \right\rangle
- \left\langle (\Delta\sigma^2)^2 \right\rangle
\qquad 
\end{eqnarray}
in which
\begin{equation}
\Delta\sigma^\ell \equiv \sigma^\ell - \left\langle \sigma^\ell \right\rangle
\quad \ell=1,2
\end{equation}
This suggests that minimizing the term $\left\langle (\sigma^1 - \sigma^2)^2 \right\rangle$
drives the action of measurement (system and apparatus evolve to states with
the same eigenvalue)
and the other two terms drive wavefunction collapse (until each system ultimately has only
a single eigenvalue $\sigma^\ell = \left\langle \sigma^\ell \right\rangle$).
However, we will find that the $\delta$-function form of $f$ is unsuitable for our objectives,
so the physical interpretation of $R^I$ is more subtle.

Next we expand in normal modes according to (\ref{C_jk_expansion_alt})
and use the eigenvalue relation (\ref{eigenvalue}):
\begin{eqnarray}
R^I &=& \nu \int^{t_f}_{t_i} \dif t_1 \, \int^{t_f}_{t_i} \dif t_2 \,
f(t_1 - t_2)
\sum_{\substack{j,k,\ell,m,\\n,p,q,r}}
\bra{\psi_j^1(t_1)} \bra{\psi_k^2(t_1)} C^*_{jk}(t_1)
\bra{\psi_\ell^1(t_2)}' \bra{\psi_m^2(t_2)}'
\nonumber\\
&&
C^*_{\ell m}(t_2) \,
\Delta_{qp} \, \Delta_{nr} \,
C_{np}(t_1) \ket{\psi_n^1(t_1)}' \ket{\psi_p^2(t_1)}' C_{qr}(t_2) \ket{\psi_q^1(t_2)} \ket{\psi_r^2(t_2)}
\end{eqnarray}

Then, using (\ref{modal_inner_product}) and defining $E_{jk} \equiv E_j^1 + E_k^2$,
\begin{eqnarray}
\label{nonlinear_interaction_C}
R^I &=&
\int^{t_f}_{t_i} \dif t_1 \, \int^{t_f}_{t_i} \dif t_2 \,
r^I(t_1,t_2)
\nonumber\\
&=& \,
\frac{1}{2}
\int^{t_f}_{t_i} \dif t_1 \, \int^{t_f}_{t_i} \dif t_2 \,
\left[ r^I(t_1,t_2) + r^I(t_2,t_1) \right]
\end{eqnarray}
where
\begin{equation}
r^I(t_1,t_2) \equiv \,
\nu
f(t_1 - t_2)
\sum_{j,k,\ell,m}
\Delta_{jm} \, \Delta_{\ell k} \,
C^*_{jk}(t_1) \, C^*_{\ell m}(t_2) \, C_{\ell m}(t_1) \, C_{jk}(t_2) \,
\me^{-\frac{\mi}{\hbar} (E_{jk}-E_{\ell m}) (t_2-t_1)}
\end{equation}
In the second line of (\ref{nonlinear_interaction_C}) we have replaced the integrand
by its real part, for the reasons discussed above, utilizing the property
\begin{equation}
\left[ r^I(t_1,t_2) \right] ^* = r^I(t_2,t_1)
\end{equation}

\subsection{Complete action and variational analysis}
Then the full action is
\begin{eqnarray}
\label{full_action}
S &=& S^1 + S^2 + S^I + R^I
\nonumber\\
&=& \int^{t_f}_{t_i} \dif t
\left[
s^{12}(t) + s^I(t)
\right]
+ \frac{1}{2} \int^{t_f}_{t_i} \dif t_1 \int^{t_f}_{t_i} \dif t_2
\left[ r^I(t_1,t_2) + r^I(t_2,t_1) \right]
\nonumber\\
&=&
\int^{t_f}_{t_i} \dif t_1 \int^{t_f}_{t_i} \dif t_2
\left\{
\frac{1}{2T}
\left[
s^{12}(t_1) + s^I(t_1)
+
s^{12}(t_2) + s^I(t_2)
\right]
+ \frac{1}{2}
\left[ r^I(t_1,t_2) + r^I(t_2,t_1) \right]
\right\}
\qquad 
\end{eqnarray}
where $s^{12}$ and $s^I$ are the integrands (including prefactors) in $(S^1 + S^2)$ and $S^I$,
as given in (\ref{action_single_system_terms_simplified}) and (\ref{interaction_local}):
\begin{equation}
s^{12} = \, B \sum_{j,k}
\left[
\lvert \dot{C}_{jk} \rvert^2
+ \frac{\mi}{\hbar} E_{jk}
\left( C^*_{jk} \,\dot{C}_{jk} - \dot{C}^*_{jk} \, C_{jk} \right)
\right]
\end{equation}
\begin{equation}
s^I =  \, \mu \sum_{j,k}
\, \Delta_{jk}^2 \, \lvert C_{jk} \rvert ^2
\end{equation}
We observe that in this form, the integrand of $S$ is real and symmetric in $t_1$ and $t_2$.
It depends on the coefficients $\{C_{pq}\}$
at two times. 
We need to find a critical point of the action
subject to the constraint (\ref{normalization_Cjk}).
In the Appendix we outline the analysis of such a problem, including the use of a
Lagrange mulitipler $\lambda(t)$ to enforce the constraint,
leading to integral equation (\ref{necessary_1_Lagrange}).
Varying $C_{jk}^*$ by that procedure and defining the operator
\begin{equation}
W \equiv \frac{\partial}{\partial C_{jk}^*(t_1)}
- \left. \frac{\partial}{\partial t_1} \right |_{t_2} \frac{\partial}{\partial \dot{C}_{jk}^*(t_1)}
\end{equation}
we find
\begin{eqnarray}
0 &=& \frac{1}{2} \, W \, s^{12}(t_1)
+ \frac{1}{2} \, W \, s^I(t_1)
+ \frac{1}{2} \int^{t_f}_{t_i} \dif t_2 \, 
W \left[ r^I(t_1,t_2) + r^I(t_2,t_1) \right]
\nonumber\\
&& + \, T \, \lambda(t_1) \,
\frac{\partial}{\partial C_{jk}^*(t_1)}
\left[ \sum_{j,k} \, \lvert C_{jk}(t_1) \rvert^2 - 1\right]
\end{eqnarray}
This becomes
\begin{equation}
\label{Euler_equation}
\ddot{C}_{jk}(t) =
\frac{2\mi}{\hbar} E_{jk} \, \dot{C}_{jk}(t)
+ \frac{1}{B} \left[ \mu \, \Delta_{jk}^2 
+ 2T \, \lambda(t)
\right] C_{jk}(t)
+ \frac{\nu}{B} \, \tilde{C}_{jk}(t)
\end{equation}
in which we define the function
\begin{equation}
\label{C_jk_tilde}
\tilde{C}_{jk}(t) \equiv 
\sum_{\ell,m} 
\Delta_{jm} \, \Delta_{\ell k} \, C_{\ell m}(t) \,
\int^{t_f}_{t_i} \dif t' \,
C^*_{\ell m}(t') \,
C_{jk}(t') \,
f(t - t') \,
\me^{-\frac{\mi}{\hbar} (E_{jk}-E_{\ell m}) (t'-t)}
\end{equation}
It can be seen by varying the action by $C_{jk}$ instead of $C_{jk}^*$
that $\lambda(t)$ must be real. 
To find it, we note that
the second derivative of the normalization condition
(\ref{normalization_Cjk}) is 
\begin{equation}
\label{2nd_derivative_normalization}
2 \sum_{j,k} \left( \lvert \dot{C}_{jk} \rvert^2 
+ \mathrm{Re} \left\{ C_{jk} ^* \, \ddot{C}_{jk} \right \}\right) = 0
\end{equation}
We eliminate $\ddot{C}_{jk}$ between (\ref{Euler_equation})
and (\ref{2nd_derivative_normalization})
and then solve for (a constant times) $\lambda(t)$:
\begin{equation}
\label{lambda}
\frac{2T \lambda}{B}
= - \sum_{j,k} \left(
\lvert \dot{C}_{jk} \rvert^2 
+ \frac{\mu}{B} \, \Delta_{jk}^2 \, \lvert C_{jk} \rvert^2
+ \mathrm{Re} \left\{
\frac{2 \mi}{\hbar} E_{jk} \, C_{jk}^* \, \dot{C}_{jk} 
+ \frac{\nu}{B} \, C_{jk}^* \, \tilde{C}_{jk} \right\}
\right)
\end{equation}
Substituting that expression into (\ref{Euler_equation}),
\begin{eqnarray}
\label{Cjk_dotdot_eqn}
\ddot{C}_{jk}
&=&
\frac{2 \mi}{\hbar} E_{jk} \, \dot{C}_{jk} 
+ \frac{\mu}{B} \, C_{jk} \left( \Delta_{jk}^2
- \sum_{\ell,m} \Delta_{\ell m}^2 \lvert C_{\ell m} \rvert ^2 \right)
\nonumber\\
&&
+ \frac{\nu}{B} \left(
\tilde{C}_{jk}
- C_{jk} \, \mathrm{Re} \left\{\sum_{\ell,m}
C_{\ell m}^* \tilde{C}_{\ell m} \right\}
\right)
- C_{jk} \sum_{\ell,m} 
\left(
\lvert \dot{C}_{\ell m} \rvert ^2 +
\mathrm{Re} \left \{ \frac{2 \mi}{\hbar} E_{\ell m} \, C_{\ell m}^* \, \dot{C}_{\ell m} \right \}
\right)
\qquad 
\end{eqnarray}
This is the equation that we expect describes the evolution of the complete
system (that is, system $+$ apparatus), as described by the coefficients
$\{C_{jk}(t)\}$ in the normal--mode expansion (\ref{C_j_expansion}).

The BCs to be applied with (\ref{Cjk_dotdot_eqn}) are specified values
of $\{C_{jk}(t_i)\}$ from the initial preparation, and the NBC at $t_f$,
which takes the form
\begin{equation}
0 = \int_{a}^{b} \dif t_2
\left. \left( \frac{\partial L}{\partial \dot{C}^*_{jk}(t_1)} \right) \right\rvert_{t_1=t_f}
\end{equation}
in which $L$ is the integrand in the full action on the last line of \eqref{full_action}.
This form of the NBC
is derived in the appendix.
\subsection{Alternative treatment of the normalization condition}
\label{Alternative_normalization}
Comparison of \eqref{Euler_equation} with \eqref{Cjk_dotdot_eqn} shows that
rigorous enforcement of the normalization condition \eqref{normalization_psi}
or \eqref{normalization_Cjk} has complicated the mathematics. Since we hope
to show that experimental results of great simplicity and generality (e.g. Born's
rule) follow from this theory, we are suspicious of the additional complexity and
wonder whether it is absolutely necessary to satisfy the stated normalization condition
at every instant $t$.

Our skepticism about that requirement is also based on a thought experiment
described by Renninger\cite{Renninger_Gedanken_Ger, *Renninger_Gedanken_Eng},
which is equivalent to the following description.
An excited atom at the origin is known to emit a photon at $t=0$, but the direction is
unknown, so the photon's wavefunction satisfies $\lvert \psi \rvert ^2 = \delta(r-ct)/(4\pi r^2)$.
A perfectly collecting hemispherical detector screen occupies the upper half
of the sphere $r=1$ light-second. Therefore, if the photon's emission direction is within
$\theta < \pi/2$, it is collected and extinguished at $t=1$ second. Otherwise, it is not
registered by the detector screen, and its wavefunction changes
to satisfy $\lvert \psi \rvert ^2 = \delta(r-ct) H(\theta-\pi/2)/(2\pi r^2)$, where $H$ is the Heaviside
function. The instantaneous change in the denominator from $4\pi r^2$ to $2\pi r^2$
at $t=1$ is not due to any measurement,
for there is none, nor to any physical change in the photon; it arises entirely from the
normalization requirement. This seems unphysical, and our suspicion deepens when
we consider that this description depends on choice of reference frame; for instance, in
any other frame the detector screen would not be (hemi)spherical but spheroidal, and
so the resulting change in magnitude of the uncollected wavefunction would happen over a
nonzero interval of time.

A more physically sound description would be that a photon intercepted by the detector
screen does not simply vanish; it interacts with (a) particle(s) of the screen to produce some
physical effect, for instance dislodging a photoelectron. A more complete
description of the experiment would include that effect. Since half of the outgoing spherical
photon wavefunction participates in that effect, it is unreasonable for the uncollected half to
double its weight to satisfy a normalization condition. We argue instead that the outgoing
uncollected photon wavefunction after $t=1$ should be normalized to integrate to $1/2$,
and with that change we see that a discontinuous and unphysical change is no longer needed
in that uncollected part at $t=1$.

Armed with our reasoning that the normalization condition \eqref{normalization_psi} is
not absolute, we propose to relax it for the experiment that is the subject of this paper.
Although for many experiments we do not expect to lose any of the wavefunction
weight in mid--experiment,
we point out that the total weight of the wavefunction (unity, meaning one particle of
whatever type is being described) is known only at $t_i$ and $t_f$. There is not, nor
can there be, any experimental evidence for a unity (or any other) value of the weight
at intermediate times. Therefore we propose that \eqref{normalization_psi} is a constraint 
only at $t_i$ and $t_f$. This is easily handled mathematically; we simply stipulate that
\eqref{normalization_Cjk} is part of the initial and final conditions.\footnote{Note of course
that if the experiment being described is the Renninger experiment, or some other
experiment with sources or sinks of the wavefunction (the quantum field),
then the normalization values at $t_i$ and $t_f$ will be modified in the ways just described,
or in more complicated ways. For instance, if the Renninger experiment were
augmented with a lower--hemisphere detector at $r=2$, then there would be
one final condition at $t=1$ and $r=1, \, \theta \le \pi/2$
and another at $t=2$ and $r=2, \, \theta \ge \pi/2$, with a normalization value of $1$
applied to the union of both collector surfaces at their respective collecting times.}
Then we can dispense
with the Lagrange multiplier altogether, so the IDE to be satisfied is
\begin{equation}
\label{Euler_equation_unconstrained}
\ddot{C}_{jk}(t) =
\frac{2\mi}{\hbar} E_{jk} \, \dot{C}_{jk}(t)
+ \frac{\mu}{B} \Delta_{jk}^2 \, C_{jk}(t)
+ \frac{\nu}{B} \, \tilde{C}_{jk}(t)
\end{equation}

\noindent
Since we regard the simplicity of this equation in comparison to \eqref{Cjk_dotdot_eqn}
as an argument for its plausibility, we will adopt it rather than the latter in the remaining
sections of the paper; nevertheless, much of the following reasoning can be applied
to \eqref{Cjk_dotdot_eqn} as well at the cost of more algebra.

\section{Comparison to desired properties}
\subsection{Stability of a superposition in the absence of a measurement}
We observe at this point that \eqref{Euler_equation_unconstrained} predicts the stability of an
unperturbed superposition, as it should. When there is no interaction
between the system and the measurement apparatus,
$\mu = \nu = 0$. The resulting equation
\begin{equation}
\label{Cjk_dotdot_eqn_no_interaction}
\ddot{C}_{jk} = \frac{2 \mi}{\hbar} E_{jk} \, \dot{C}_{jk} 
\end{equation}
has the solution $\dot{C}_{jk} =0 \; \forall j, \! k$, that is, stability of the superposition.
Furthermore, since (for each subsystem $\ell=1$ or $2$)
the modes in the expansion (\ref{C_jk_expansion_alt})
were defined as solutions of the no--measurement wave equation, the stable solution
resulting from our analysis here agrees with the solution of the ordinary wave equation
for each isolated system.
\subsection{Collapse to a single eigenstate with $\sigma_j^1 = \sigma_k^2$}
\label{Collapse_single_eigenstate}
This includes three events we expect in a measurement: system 1 must
collapse to a single eigenstate of $\sigma_{op}^1$, or a superposition of
eigenstates with the same eigenvalue; system 2 must similarly collapse;
and the eigenvalues of the two systems must agree.
The third condition (measurement) requires that for any $j,k$,
\begin{equation}
\label{measurement_condition}
\Delta_{jk}=0 \quad \text{or} \quad C_{jk}=0
\end{equation}
Although we will not analyze the differential equation
\eqref{Euler_equation_unconstrained}
to describe the approach
to these three conditions,
we will show that it is consistent with their satisfaction
in the steady state, when all time
derivatives of $\{C_{pq}\}$ vanish.
Thus it is plausible for the combined system to reach such a state,
and having done so, to remain in that state.

We see that condition (\ref{measurement_condition})
together with the steady--state condition cause every term in
(\ref{Euler_equation_unconstrained}) to vanish except possibly 
the last. To understand those terms, consider that after the system attains
a steady state, we can replace all the factors $C_{pq}$ or $C^*_{pq}$
on the RHS of (\ref{C_jk_tilde}) by their final values, which satisfy
(\ref{measurement_condition}). Then at times $t$ greater than $\tau$
after the full system reaches its steady state, any nonzero terms $\ell,m$
on the RHS must have
\begin{equation}
\label{Deltas_0}
\Delta_{jk} = \Delta_{\ell m} = 0
\end{equation}
If either of
systems 1 and 2 has collapsed to a single state (or a set of states with
a single eigenvalue), then by (\ref{measurement_condition}) the other
system has also collapsed, and it is easy to see that (\ref{Deltas_0})
implies that $\Delta_{jm} = \Delta_{\ell k} = 0$, so the only possible nonzero
term in $\tilde{C}_{jk}$ is zero after all. Therefore the last term in
\eqref{Euler_equation_unconstrained} vanishes, so the equation is consistent
with the supposed late--time steady state. On the other hand, if systems
1 and 2 have not collapsed, there are terms in (\ref{C_jk_tilde}) that do
not trivially vanish. We conclude that the evolution equation predicts that
a late--time steady state is only possible if both the measurement condition
is satisfied (the apparatus state corresponds to the state of the system
being measured) and both systems have collapsed to a single eigenvalue.

We would prefer to have a more rigorous analysis, both disposing of the
possibility that the combined system never reaches a steady state and
describing the approach to the steady state. This analysis must await
future work, possibly including numerical studies. Our objective in this
paper is to show the possibility that a variational principle of the type
we have developed can explain the measurement problem.
\subsection{Consistency with Born's rule}
\label{Borns_rule}
The well--known experimental observation is that in an ensemble
of identically--prepared measurements
of some property (eigenvalue), beginning with a system
in a superposition of modes with different values
of the eigenvalue, the expected proportion of outcomes equal to a particular value
will be the the weight of that value in the superposition.
(At this point we take it as given that the system will
collapse to a single value of the eigenvalue.)
In our case, where the system being measured
is denoted $\ell=1$, the weight corresponding to eigenvalue
$\sigma^1_j$ is
\begin{equation}
\label{P_j}
P_j \equiv \sum_k \lvert C_{jk}(t_i) \rvert^2
\end{equation}
(More generally, it is $\sum_{j,k} \lvert C_{jk}(t_i) \rvert^2$, where
the sum on $j$ is over all modes with a single value of the eigenvalue.
For simplicity, we will consider only the non--degenerate case, but the
extension to the more general case should be straightforward.)

It will be convenient to denote averages over an ensemble of
identically prepared
experimental realizations by an overbar. Then, if it is taken as given
that the collapse to a single eigenvalue is complete by $t_f$,
we can see that the relation
\begin{equation}
\label{equality_of_averaged_probabilities}
\overline{P_j(t_i)} = \overline{P_j(t_f)} \quad \forall j
\end{equation}
is equivalent to Born's rule.
This equivalence holds because at the initial time $t_i$,
by the requirement of identical preparation, every member of the
ensemble contributes the same value
$P_j(t_i)$ to the ensemble average.
At $t_f$, $P_j = 1$ in a fraction $P_j(t_i)$ of the realizations in the
ensemble, and 0 in the others.
So (\ref{equality_of_averaged_probabilities}) is the relation that
should be predicted by a successful theory.

We would like to be able prove that Born's rule
(\ref{equality_of_averaged_probabilities})
follows from our nonlocal wave equation (\ref{Cjk_dotdot_eqn}).
The theoretical proof has eluded us so far; we may ultimately have to rely
on numerical studies. However, we sketch out here some of the ideas
that may contribute to the theoretical analysis.

By differentiating (\ref{P_j}) twice, we see that (supposing
that by the system preparation $\dot{P}_j(t_i)=0$)
\begin{eqnarray}
\label{P_j_dot}
\dot{P}_j(t) &=& 2 \int_{t_i}^t dt' \, \sum_k \left( \lvert \dot{C}_{jk} \rvert^2 
+ \mathrm{Re} \left\{ C_{jk} ^* \, \ddot{C}_{jk} \right \}\right)
\nonumber\\
&=& 2 \int_{t_i}^t dt' \, \sum_k \left[
\lvert \dot{C}_{jk} \rvert^2 
+ \frac{\mu}{B} \, \Delta_{jk}^2 \lvert C_{jk} \rvert^2
+ \mathrm{Re} \left\{
\frac{2 \mi}{\hbar} E_{jk} \, C_{jk}^* \, \dot{C}_{jk} 
+ \frac{\nu}{B} \, C_{jk}^* \, \tilde{C}_{jk} \right\}
\right.
\end{eqnarray}
In the term in the integrand involving $E_{jk}$,
let $C_{jk} = X \me^{\mi \phi}$ for real $X$ and $\phi$.
Then
\begin{eqnarray}
\mathrm{Re} \, \left\{\frac{2 \mi}{\hbar} E_{jk} \, C_{jk}^* \, \dot{C}_{jk} \right\} 
&=& \mathrm{Re} \, \left\{\frac{2 \mi}{\hbar} E_{jk} X (\dot{X} + \mi X \dot{\phi}) \right\}
\nonumber\\
&=& -\frac{2}{\hbar} E_{jk} X^2 \dot{\phi}
\end{eqnarray}
so
\begin{equation}
\overline{ \mathrm{Re} \, \left\{\frac{2 \mi}{\hbar} E_{jk} \, C_{jk}^* \, \dot{C}_{jk} \right\} }= 0
\end{equation}
by symmetry, since the phase $\phi$ is equally likely to increase or decrease.

To deal with the term $\mathrm{Re} \, \{\frac{\nu}{B}  \, C^*_{jk} \, \tilde{C}_{jk} \}$, we note that
\begin{eqnarray}
\label{Cjkstar_Cjktilde}
C_{jk}^* (t) \, \tilde{C}_{jk} (t)
&=& \,
\Delta_{jk}^2 \,
\langle \langle C_{jk}^2(t) \rangle \rangle \,
\lvert C_{jk}(t) \rvert ^2
\nonumber\\
&+& C_{jk}^* (t) \, \sideset{}{'}\sum_{\ell,m}
\Delta_{jm} \, \Delta_{\ell k} \, C_{\ell m}(t)
\int^{t_f}_{t_i} \dif t' \,
C^*_{\ell m}(t') \, C_{jk}(t') \,
f(t - t') \,
\me^{-\frac{\mi}{\hbar} (E_{jk}-E_{\ell m}) (t'-t)}
\end{eqnarray}
in which we define the ``moving average''
\begin{equation}
\langle \langle C_{jk}^2(t) \rangle \rangle \equiv
\int^{t_f}_{t_i} \dif t' \, f(t - t') \,\lvert C_{jk}(t') \rvert ^2
\end{equation}
and the primed sum denotes the sum over all $\ell, m$
except the single term $\ell=j, m=k$.

We have hypothesized that the solution $\{C_{jk}(t) \,\, \forall j,k,t \}$
of the variational principle is constrained by the initial (preparation)
condition at $t_i$ and the final (NBC) condition at $t_f$.
We now venture a little further and suppose that the desired solution,
in order to extremize the action, uses the entire interval from
$t_i$ to $t_f$ to evolve from initial to final values of $\{C_{jk}\}$;
this is plausible due to the term in the action 
[$\braket{\dot{\psi}^\ell}{\dot{\psi}^\ell}$
in (\ref{Ll_form}) or 
$\int \dif t \lvert \dot{C}_{jk} \rvert^2$
in (\ref{action_single_system_terms_simplified})]
that penalizes rapid
transitions.
Therefore $\lvert \dot{C}_{jk} \rvert \sim 1/T$.
But a measurement adequate to resolve two states $j,k$ and $\ell,m$
with $E_{jk} \neq E_{\ell m}$ is conventionally understood to require a duration
\begin{equation}
T \gg \frac{\hbar}{\lvert E_{jk} - E_{\ell m} \rvert}
\end{equation}
We conclude therefore that there is an $\epsilon$ such that
\begin{equation}
\frac{\hbar \lvert \dot{C}_{pq} \rvert}{\lvert E_{jk} - E_{\ell m} \rvert}
 < \epsilon \ll 1
\end{equation}
for any $p,q$ and for any choice of $j,k,\ell,m$ for which $E_{jk} \neq E_{\ell m}$.
We may also require the function $f$ to be slowly varying in the sense that
\begin{equation}
\frac{\hbar \lvert \dot{f} \rvert}{\lvert E_{jk} - E_{\ell m} \rvert f_{max}}
 < \epsilon \ll 1
\end{equation}
where $f_{max}$ is the maximum value taken by $f$.
Consequently, with the additional assumption that $E_{jk} = E_{\ell m}$
only if $j=\ell$ and $k=m$, the integral in the second term of
(\ref{Cjkstar_Cjktilde}) can be integrated by parts twice:
\begin{eqnarray}
\label{integral_from_Cjk_tilde}
\int^{t_f}_{t_i} & \dif t' \, &
C^*_{\ell m}(t') \, C_{jk}(t') \,
f(t - t') \,
\me^{-\frac{\mi}{\hbar} (E_{jk}-E_{\ell m}) (t'-t)}
\nonumber\\
&=&
\, \frac{\mi\hbar}{E_{jk}-E_{\ell m}}
\left[
C^*_{\ell m}(t') \, C_{jk}(t') \,
f(t - t') \,
\me^{-\frac{\mi}{\hbar} (E_{jk}-E_{\ell m}) (t'-t)}
\right]_{t'=t_i}^{t_f}
\nonumber\\
&&+ \, \frac{\hbar^2}{(E_{jk}-E_{\ell m})^2}
\left\{
\frac{\dif}{\dif t'} \left[ C^*_{\ell m}(t') \, C_{jk}(t') \, f(t - t') \right]
\me^{-\frac{\mi}{\hbar} (E_{jk}-E_{\ell m}) (t'-t)}
\right\}_{t'=t_i}^{t_f}
\nonumber\\
&&+ \, \frac{\hbar^2}{(E_{jk}-E_{\ell m})^2}
\int^{t_f}_{t_i} \dif t' \,
\frac{\dif^2}{\dif t'\,^2} \left[ C^*_{\ell m}(t') \, C_{jk}(t') \, f(t - t') \right]
\me^{-\frac{\mi}{\hbar} (E_{jk}-E_{\ell m}) (t'-t)}
\nonumber\\
&=&
\, \frac{\mi\hbar}{E_{jk}-E_{\ell m}}
\left[
C^*_{\ell m}(t') \, C_{jk}(t') \, f(t - t') \,
\me^{-\frac{\mi}{\hbar} (E_{jk}-E_{\ell m}) (t'-t)}
\right]_{t'=t_i}^{t_f}
[1 + O(\epsilon)]
\end{eqnarray}
Our hypothesis to explain the apparent randomness of quantum mechanical measurements
is that some ``hidden variable'' is not sufficiently well controlled in typical practice to 
determine a single outcome. Here the hidden variable appears to be the stop time $t_f$
or equivalently the duration $T$ of the experiment.
If the uncertainty in $t_f$ is $\gg 1/\Delta E$ for the smallest energy difference $\Delta E$,
the realization average of the complex exponential factor
$\mathrm{exp}[-\frac{\mi}{\hbar} (E_{jk}-E_{\ell m}) (t_f-t)]$
is zero. We would like to infer from that, neglecting $O(\epsilon)$,
that the realization average of
(\ref{integral_from_Cjk_tilde}) vanishes, but there are two problems.
We cannot factor the realization average
\begin{equation}
\overline{
C^*_{\ell m}(t_f) \, C_{jk}(t_f) \,
\me^{-\frac{\mi}{\hbar} (E_{jk}-E_{\ell m}) (t_f-t)}
}
\ne
\overline{
C^*_{\ell m}(t_f) \, C_{jk}(t_f)
} \,\,
\overline{
\me^{-\frac{\mi}{\hbar} (E_{jk}-E_{\ell m}) (t_f-t)}
}
\end{equation}
because the final values of the coefficients $C^*_{\ell m}$ and $C_{jk}$
are correlated with the complex exponential factor. Also, the $t'=t_i$
term in (\ref{integral_from_Cjk_tilde}) will not average to zero;
since the initial conditions are imposed at the start time,
uncertainty is $t_i$ is presumably not a source of variation in the outcome.

From the surviving terms in the realization average of (\ref{P_j_dot}) we see that
\begin{eqnarray}
\overline{P_j(t_f)} - \overline{P_j(t_i)} &=& \int_{t_i}^{t_f} \dif t' \, \overline{\dot{P}_j(t')}
\nonumber\\
&=& 2 \int_{t_i}^{t_f} \dif t' \int_{t_i}^{t'} \dif t'' \,
p(t'')
\nonumber\\
&=& 2 \int_{t_i}^{t_f} \dif t'' \, (t_f - t'') \,
p(t'')
\end{eqnarray}
and therefore
\begin{equation}
\label{probability_difference}
\left\lvert \overline{P_j(t_f)} - \overline{P_j(t_i)} \right\rvert
< 2 \, T \int_{t_i}^{t_f} \dif t'' \,
\left\lvert
p(t'')
\right\rvert
\end{equation}
with
\begin{equation}
p \equiv
 \sum_k
\overline{
\lvert \dot{C}_{jk} \rvert ^2
+ \frac{\mu}{B} \, \Delta_{jk}^2 \lvert C_{jk} \rvert^2
+ \frac{\nu}{B} \, \Delta_{jk}^2 \, 
\langle \langle C_{jk}^2 \rangle \rangle \,
\lvert C_{jk} \rvert ^2
}
\end{equation}
If the previously identified issues in the proof of Born's rule are resolved,
it remains to show that the LHS of (\ref{probability_difference})
vanishes, at least in the limit at $T\rightarrow \infty$. 
(As noted earlier, experimental results at variance with Born's rule are likely to
be rejected as invalid if $T$ is too small.)
To do that, we must show that $p(t)$ decays fast enough
that the integral in (\ref{probability_difference})
decreases faster than $1/T$.
\section{Discussion}
\subsection{Sensitivity of the system evolution to a measurement}
Traditional discussions of quantum mechanics maintain that making a measurement
changes the evolution of a quantum system from its unitary evolution, as described
by the wave equation, to a collapsed state, as described by the measurement side
of the bipartite theory. Thus the unitary evolution cannot be observed without
interrupting it. This remarkable sensitivity to observation is not explained except
as the inevitable corollary of the special treatment of measurement in the theory.

We also find this sensitivity to observation in our picture, but can give more of an
explanation for it. The act of measuring a system involves causing it to physically
interact with a measurement apparatus, and the variational principle describes
the evolution of the combined system. The readout of the measurement at $t_f$
defines the end of the domain of integration of the variational principle. Of
course, the theory continues to apply after $t_f$, but the observation
at $t_f$, like its preparation at $t_i$ and its spatial boundary
conditions, imposes a leakproof barrier
to influences from outside the problem domain, so that a solution may be
found within that domain without reference to the rest of the universe.

Now if the measurement apparatus were read at some intermediate time $t_m$,
the structure of the problem would be different. Instead of applying between
$t_i$ and $t_f$, the variational principle would apply twice, from
$t_i$ to $t_m$ and from $t_m$ to $t_f$. The appearance of
a constraint at $t_m$ as a final condition on the first interval
and an initial condition on the second would make this a different
problem than the original one from $t_i$ to $t_f$. (As we have
explained, the intervention at $t_m$ results in the appearance
of an NBC on the solution between $t_i$ and $t_m$,
even though it does not dictate the result of the reading at $t_m$.)
Consequently, the act of observing the system at $t_m$ changes it,
just as in conventional interpretations.

The reader may object that we have not removed the mystery but moved it
to a different concept. Instead of declaring by fiat that a measurement changes
the system, we have declared that the domain of integration of the variational
principle must end at the time (and place) at which the measurement apparatus is read.
We haven't explained what is special about the 
events at $t_f$ that allow us to end the domain there.

The criticism is valid, but we point out that we have pushed back the mystery, or made
it less mysterious, by relating it to considerations of BCs. Certainly the description of
a measurement in terms of an action integral bounded at $t_i$ and $t_f$ must be an
approximation to a more complete theory that includes a greater time interval before
and after $[t_i,t_f]$ and a fuller description of the measurement process.
On the other hand, the empirical
fact that broad statements of great generality apply to measurements, regardless of 
the system under study or the mechanism of the process, strongly suggests that a
simple description is possible, particularly regarding a time before the measurement
($t_i$) and a time after its completion ($t_f$). The validity of the simple description
is not necessarily a surprise; it may be that the interactions that can be so described
have been adopted as measurement procedures precisely because of their ability to
give repeatable quantitative results.

If the simple description proposed in this paper
turns out to be successful in description and prediction at some level of
approximation, that will be evidence of its usefulness, without denying the possibility
of a more complete theory. Eventually such an improved theory may show
that collapse/decay to a single eigenvalue occurs
at $t_f$ in a physically justifiable way, based on the role of the apparatus in the action,
and so it is appropriate to simplify the problem as we have done
by terminating the integral at $t_f$ and accepting the NBC there.

An extended analysis of that type would also be appropriate
to explore another aspect of the new theory.
We have argued that we can solve the variational principle between
$t_i$ and $t_f$, which would presumably enable a prediction of the experimental outcome at
$t_f$ (based on (a) fixed value(s) of hidden variable(s), of course). We have asserted that the
final condition at $t_f$ provides a leakproof barrier to influences from outside that problem domain.
But the theory must apply under reversal of the direction of time, so it should also be possible
to apply an experimental preparation (initial condition) at $t_{f2} \equiv t_f+T$
and a measurement readout
(NBC as a final condition) at $t_f$ to predict an outcome at $t_f$ based
on physics between $t_f$ and $t_{f2}$.
We suspect that the theory retains sufficient flexibility to allow the two solutions (for
$t_i \le t \le t_f$ and $t_f \le t \le t_{f2}$) to agree at $t_f$. It probably helps that we expect
(in both cases) to apply \emph{natural} BCs at $t_f$, so we are not
actually constraining the value of the measured variable. Also, continuity constraints
on fields, wavefunctions and derivatives appearing in the action may help to avoid
contradictions.
Since these two predictions must agree, the barrier
at $t_f$ is not completely leakproof.
It is rather a partially permeable membrane,
as suggested by the applicability of an NBC that constrains some
but not all properties of the system at $t_f$. 
This type of study may give insight
into the nature of the constraint imposed by the measurement readout.
\subsection{Causality and time--ordering issues}
Retrocausality---the dependence of phenomena at a given time on phenomena in
their future---conflicts with the usual notion of causality---the concept that causes
precede their effects in time. However, multiple
authors\cite{Cramer_1986, Price_1996, Schulman_1997} have pointed
out that such a notion of causality is not necessary to avoid contradictions. If event
$A\Rightarrow B$, then $B \Rightarrow \, \sim \! A$ would produce a contradiction.
But if we are somehow prevented from declaring that $B \Rightarrow \, \sim \! A$
(or an equivalent combination of statements), then in principle
$A\Rightarrow B$ is possible \emph{even if $B$ occurs earlier than $A$}.

To apply this to our use of retrocausality in the variational principle,
we are asserting that the NBC at $t_f$
(which applies because a measurement is made at that time,
even though the result of the measurement is unconstrained)
is an event $A$ that constrains the solution between $t_i$
and $t_f$, so that solution at some intermediate time $t_m$
can be considered as event $B$.
But the event $B$ thus chosen is by definition consistent with $A$,
since it is a point along the solution based on $A$. It is
not possible to claim that $B \Rightarrow \, \sim \! A$, so
no contradiction is possible.

Of course, the usual objection to this is that one could
intervene at $t_m$ to change the trajectory of events and
produce $\sim \! A$ at $t_f$ (going back in time and shooting one's
grandparent, in the usual cliche). But doing this changes the
problem, as described above; now the variational principle
applies from
$t_i$ to $t_m$ and from $t_m$ to $t_f$, with the
intervention imposing new BCs at $t_m$.
Since this is a different
problem than the original one, the
original solution does not apply and no claim of a
contradiction can be made.
\subsection{Choice of the function $f$}
We have relied on a supposed interaction between wavefunctions at $t_1$ and $t_2$,
as expressed in the nonlocal action term (\ref{first_nonlinear_interaction}).
The interaction is a \emph{physical} process with a temporal range
described by the function $f$. It will be important to determine the form of $f$;
this may be explored numerically, but additional physical insight could be very
useful.

Our earlier hypotheses that $f$ is a decreasing function of the absolute value of
its argument and that it has a finite range $\tau$ are intuitively appealing, but
they are not the only possibility. In fact, we cannot rule out the opposite extreme,
that $f(t) \equiv 1$. This would mean that the nonlocal interaction has infinite
range, but in practice for a given measurement it would be limited to the
interval $[t_i,t_f]$.
(Without the finite--range limit $\tau$, our analysis in section
\ref{Collapse_single_eigenstate} would have to be revisited.)
\subsection{Solving the integrodifferential equation}
As mentioned above, it will be important to solve, or otherwise study,
the IDE (\ref{Euler_equation_unconstrained}). That effort may be made theoretically,
or numerically if need be. We would like to understand under what
conditions the system reaches the collapsed state described in
section \ref{Collapse_single_eigenstate}, how fast that late--time
state is approached, and which of the possible collapsed states
is reached, as a function of the hidden variable(s).
It will also be important to test whether the equation produces
outcome frequencies consistent with Born's rule, possibly
following ideas in section \ref{Borns_rule}.

One question is whether, given a choice of initial conditions and
hidden variable(s), the solution to the IDE is unique (and even
whether a solution exists). If there is always a unique solution,
the theory may be completely deterministic (although it remains
to be seen what that means for a retrocausal theory), so we may
be able to dispense completely with the idea that quantum
mechanical processes depend on \emph{instrinsically random}
variables. Such a discovery might have far--reaching
ramifications in quantum information technologies that rely
on (supposed) randomness.

If this understanding enables us to make predictions based on
the theory, we will look for experimentally testable predictions.
Although we have argued that the new theory will agree with
many features of conventional theory, it is certainly possible
that it could differ in some ways.\footnote{We expect that it will differ
in the normalization factor applied to the wavefunction in
experiments like that of Renninger, as discussed above,
but that is a difference in how a physical state is described
mathematically, not a difference in the state itself, and so not
experimentally testable.} One possibility is that results
that have historically been seen to vary, supposedly due to 
intrinsic randomness, may vary less or not at all
if a hidden (that is, historically uncontrolled) variable is 
controlled in new experiments (guided by 
new predictions about how well or to what values it must
be controlled).

Of course, it is possible that the particular choice of action we have
made, and the IDE resulting from it, do not correspond to nature.
Even in that case, our exposition here shows that a variational
principle of this type, including our assumptions of retrocausality,
nonlocality, and one or more hidden variables, can lead to a
plausible theory that avoids, resolves or explains
problematic features of conventional quantum theory.
If the theory presented here is not borne out, a
similarly--constructed theory with a different form of the action
may be more successful.
\section{Acknowledgments}
The author appreciates the support of the National Nuclear Security Agency's
Advanced Scientific Computing (ASC) program, and
useful discussions with Kenneth Wharton and Daniel
Sheehan. Most importantly, the ideas were developed
and discussed over a long period of time with Dale W. Harrison, without whom this
work would not have been possible.

\appendix*
\section{Calculus of Variations: Two--time Variant}
\label{Two_time_variant}
A basic problem in the calculus of variations \cite{Courant} is to find the function $\phi(t)$
for which the integral
\begin{equation}
\label{variational_basic}
S[\phi] = \int_{a}^{b} \dif t \, F(t,\phi(t),\dot{\phi}(t))
\end{equation}
is stationary with respect to infinitesimal changes in the function $\phi$.
Here $F$ is a given function with continuous first partial derivatives
and piecewise continuous second derivatives.
The function $\phi(t)$ is required to be continuous with piecewise continuous
first derivative, and must satisfy
\begin{equation}
\label{fixed_BCs}
\phi(a)=A \quad \phi(b)=B
\end{equation}
for given $A$ and $B$.
Under these conditions a necessary condition for (\ref{variational_basic}) is
the Euler equation
\begin{equation}
0 = \frac{\partial F}{\partial \phi}
 - \frac{\dif}{\dif t} \frac{\partial F}{\partial \dot{\phi}}
\end{equation}
\subsection{Two--time variant}
In our case the integrated function $F$ depends on the
unknown function $\phi$ at two times, both of which are integrated over:
\begin{equation}
\label{two_time_functional}
S[\phi] = \int_{a}^{b} \dif t_1 \, \int_{a}^{b} \dif t_2 \, 
F(t_1,t_2,\phi(t_1),\dot{\phi}(t_1),\phi(t_2),\dot{\phi}(t_2))
\end{equation}
As in the standard derivation, we find a necessary condition by defining
\begin{equation}
\label{necessary_epsilon}
\theta(t,\epsilon) = \phi(t) + \epsilon \, \eta(t)
\end{equation}
and requiring that
\begin{equation}
\left. \frac{\dif S[\theta]}{\dif \epsilon} \right|_{\epsilon=0} = 0
\end{equation}
for any continuous function $\eta(t)$ with piecewise continuous derivative
and
\begin{equation}
\label{eta_zeroAB}
\eta(a)=\eta(b)=0
\end{equation}
Condition (\ref{necessary_epsilon}) becomes
\begin{eqnarray}
\label{eta_integral}
0 &=& \int_{a}^{b} \dif t_1 \, \int_{a}^{b} \dif t_2 \,
\left[ \eta(t_1) \frac{\partial F}{\partial \phi(t_1)} + \dot{\eta}(t_1) \frac{\partial F}{\partial \dot{\phi}(t_1)}
+ \eta(t_2) \frac{\partial F}{\partial \phi(t_2)} + \dot{\eta}(t_2) \frac{\partial F}{\partial \dot{\phi}(t_2)} \right]
\nonumber\\
&=& \int_{a}^{b} \dif t_1 \, \int_{a}^{b} \dif t_2 \,
\left[ \eta(t_1) \left( \frac{\partial F}{\partial \phi(t_1)}
- \left. \frac{\partial}{\partial t_1} \right |_{t_2} \frac{\partial F}{\partial \dot{\phi}(t_1)} \right)
+ \eta(t_2) \left( \frac{\partial F}{\partial \phi(t_2)}
- \left. \frac{\partial}{\partial t_2} \right |_{t_1}\frac{\partial F}{\partial \dot{\phi}(t_2)} \right) \right]
\qquad 
\end{eqnarray}
Since $\eta(t)$ is arbitrary (subject to the restrictions already stated), this requires that
\begin{equation}
\label{necessary_1}
0 = \int_{a}^{b} \dif t_2 \, \left( \frac{\partial F}{\partial \phi(t_1)}
- \left. \frac{\partial}{\partial t_1} \right |_{t_2} \frac{\partial F}{\partial \dot{\phi}(t_1)} \right)
\end{equation}
and
\begin{equation}
\label{necessary_2}
0 = \int_{a}^{b} \dif t_1 \, \left( \frac{\partial F}{\partial \phi(t_2)}
- \left. \frac{\partial}{\partial t_2} \right |_{t_1}\frac{\partial F}{\partial \dot{\phi}(t_2)} \right)
\end{equation}
as necessary conditions for the stationarity of $S[\phi]$.
\subsection{Special cases}
A special case of interest is when $F$ factors into
$t_1$--dependent and $t_2$--dependent factors:
\begin{equation}
\label{F_factors}
F(t_1,t_2,\phi(t_1),\dot{\phi}(t_1),\phi(t_2),\dot{\phi}(t_2)) = G(t_1,\phi(t_1),\dot{\phi}(t_1)) H(t_2,\phi(t_2),\dot{\phi}(t_2))
\end{equation}
so that
\begin{equation}
\label{factored_integral}
S[\phi] = \int_{a}^{b} \dif t_1 \, 
G(t_1,\phi(t_1),\dot{\phi}(t_1))
\int_{a}^{b} \dif t_2 \, 
H(t_2,\phi(t_2),\dot{\phi}(t_2))
\end{equation}
and the necessary conditions (\ref{necessary_1}) and (\ref{necessary_2}) become
\begin{equation}
\label{local-stationarity_1}
0 = \frac{\partial G}{\partial \phi(t_1)}
- \frac{\dif}{\dif t_1} \frac{\partial G}{\partial \dot{\phi}(t_1)}
\end{equation}
and
\begin{equation}
\label{local-stationarity_2}
0 = \frac{\partial H}{\partial \phi(t_2)}
- \frac{\dif}{\dif t_2} \frac{\partial H}{\partial \dot{\phi}(t_2)}
\end{equation}
if we exclude the possibility that either of the integrals in (\ref{factored_integral}) vanishes.
These relations are of course the stationarity conditions for those
two integrals if they were
considered independently. We observe that the special case in which $F$ factors
as in (\ref{F_factors}) is significantly different than the general case, in that the
solution of the former can be expressed as differential equations but the latter
requires integral equations.

In this paper our concern is limited to functions $F$ that are symmetric in $t_1$
and $t_2$, that is, invariant under their interchange. For this special case,
equations (\ref{necessary_1}) and (\ref{necessary_2}) are equivalent, as
are (\ref{local-stationarity_1}) and (\ref{local-stationarity_2}).
\subsection{Natural boundary condition}
Consider the case in which the boundary conditions (\ref{fixed_BCs})
are replaced by
\begin{equation}
\label{one_fixed_BC}
\phi(a)=A
\end{equation}
that is, the solution is not constrained at $t=b$ (except, as will be shown, by the NBC).
Then condition (\ref{eta_zeroAB}) is replaced by
\begin{equation}
\label{eta_zeroA}
\eta(a)=0
\end{equation}
(no constraint on $\eta(b)$) and so the second line
of (\ref{eta_integral}) becomes
\begin{eqnarray}
0 &=& \int_{a}^{b} \dif t_1 \, \int_{a}^{b} \dif t_2 \,
\left[ \eta(t_1) \left( \frac{\partial F}{\partial \phi(t_1)}
- \left. \frac{\partial}{\partial t_1} \right |_{t_2} \frac{\partial F}{\partial \dot{\phi}(t_1)} \right)
+ \eta(t_2) \left( \frac{\partial F}{\partial \phi(t_2)}
- \left. \frac{\partial}{\partial t_2} \right |_{t_1}\frac{\partial F}{\partial \dot{\phi}(t_2)} \right) 
\right]
\nonumber\\
&&+
\int_{a}^{b} \dif t_2 \,\,
\eta(t_1) \left. \left( \frac{\partial F}{\partial \dot{\phi}(t_1)} \right) \right\rvert_{t_1=a}^b
\end{eqnarray}
But since the functions $\eta(t)$ satisfying (\ref{eta_zeroAB}) are among the set
of functions allowed by (\ref{eta_zeroA}),
$F$ must satisfy (\ref{necessary_1}) and (\ref{necessary_2}),
so the last equation becomes simply
\begin{eqnarray}
0 &=& \int_{a}^{b} \dif t_2 \,\,
\eta(t_1) \left. \left( \frac{\partial F}{\partial \dot{\phi}(t_1)} \right) \right\rvert_{t_1=a}^b
\nonumber\\
&=& - \eta(b) \int_{a}^{b} \dif t_2 \,
\left. \left( \frac{\partial F}{\partial \dot{\phi}(t_1)} \right) \right\rvert_{t_1=b}
\end{eqnarray}
so we find that the NBC is
\begin{equation}
0 = \int_{a}^{b} \dif t_2 \,\,
\left. \left( \frac{\partial F}{\partial \dot{\phi}(t_1)} \right) \right\rvert_{t_1=b}
\end{equation}
and of course by symmetry
\begin{equation}
0 = \int_{a}^{b} \dif t_1 \,\,
\left. \left( \frac{\partial F}{\partial \dot{\phi}(t_2)} \right) \right\rvert_{t_2=b}
\end{equation}
\subsection{Lagrange multipliers}
A related problem is to find a stationary point of $S[\phi]$,
as given by (\ref{two_time_functional}), subject to a constraint
\begin{equation}
\label{constraint}
K(t,\phi(t)) = 0 \qquad \forall t
\end{equation}
This can be addressed by the method of Lagrange multipliers, in a straightforward
extension of the derivation given in reference \cite{Courant}. For the special case
of symmetric $F$, that analysis shows that we can introduce a Lagrange multiplier
$\lambda(t)$ and replace condition
(\ref{necessary_1}) by
\begin{eqnarray}
\label{necessary_1_Lagrange}
0 &=& \int_{a}^{b} \dif t_2 \, \left( \frac{\partial F}{\partial \phi(t_1)}
- \left. \frac{\partial}{\partial t_1} \right |_{t_2} \frac{\partial F}{\partial \dot{\phi}(t_1)}
+ \lambda(t_1) \left. \frac{\partial K}{\partial \phi} \right|_{t=t_1}
\right)
\nonumber\\
&=& (b-a) \, \lambda(t_1) \left. \frac{\partial K}{\partial \phi} \right|_{t=t_1}
+ \int_{a}^{b} \dif t_2 \, \left( \frac{\partial F}{\partial \phi(t_1)}
- \left. \frac{\partial}{\partial t_1} \right |_{t_2} \frac{\partial F}{\partial \dot{\phi}(t_1)}
\right)
\end{eqnarray}
The solution of this differential equation is $\phi$ as a function of $t$ and the entire
function $\lambda$. Finally, $\lambda(t)$ is determined by requiring the satisfaction
of (\ref{constraint}).

\bibliography{quantum_foundations}

\end{document}